\documentclass[a4paper,onecolumn,allowtoday,accepted=2025-03-05]{quantumarticle}
\pdfoutput=1
\usepackage[T1]{fontenc}
\usepackage[british]{babel}
\usepackage[unicode,colorlinks]{hyperref}
\usepackage{graphicx}
\usepackage{amsmath,amssymb,amsthm,bm}
\usepackage{xspace}
\usepackage{mathtools}
\usepackage{dsfont}
\usepackage{xcolor}
\usepackage{tikz,pgfplots}\pgfplotsset{compat=1.18}
\usepackage{subfig}

\usepackage{csquotes}
\usepackage[backend=bibtex,style=phys,biblabel=brackets,eprint=true,doi=false,url=true,maxnames=10,defernumbers=true,giveninits=true]{biblatex}
\bibliography{biblio}

\DeclareMathOperator{\tr}{tr}

\let\originalleft\left
\let\originalright\right
\renewcommand{\left}{\mathopen{}\mathclose\bgroup\originalleft}
\renewcommand{\right}{\aftergroup\egroup\originalright}

\newcommand{\Ket}[1]{{ \vert {#1}  \rangle \!  \rangle}}

\newcommand{\bra}[1]{\left\langle #1 \right|}
\newcommand{\ket}[1]{\left| #1 \right\rangle}

\newcommand{\ketbra}[2]{\left|#1\middle\rangle\middle\langle#2\right|}
\newcommand{\proj}[1]{\left|#1\middle\rangle\middle\langle#1\right|}

\newcommand{\id}{\mathds{1}}

\newtheorem*{theorem*}{Theorem}

\mathtoolsset{centercolon}

\newcommand{\lmap}{\mathcal L}
\newcommand{\g}{\mathcal G}
\newcommand{\z}{\mathcal Z}
\newcommand{\gmap}{\widehat{\mathcal G}}
\newcommand{\zmap}{\widehat{\mathcal Z}}
\renewcommand{\vec}{\operatorname{vec}}
\newcommand{\svec}{\operatorname{svec}}
   
\hypersetup{pdftitle={Quantum key distribution rates from non-symmetric conic optimization}}

\begin{document}
\title{Quantum key distribution rates from non-symmetric conic optimization}
\author{Andrés González Lorente}\orcid{0009-0009-1689-5682}\affiliation{Departamento de Física Teórica, Atómica y Óptica, Universidad de Valladolid, 47011 Valladolid, Spain}
\author{Pablo V. Parellada}\orcid{0000-0002-6768-6671}\affiliation{Departamento de Física Teórica, Atómica y Óptica, Universidad de Valladolid, 47011 Valladolid, Spain}
\author{Miguel Castillo-Celeita}\orcid{0000-0003-2905-0389}\affiliation{Departamento de Física Teórica, Atómica y Óptica, Universidad de Valladolid, 47011 Valladolid, Spain}
\author{Mateus Araújo}\orcid{0000-0003-0155-354X}\affiliation{Departamento de Física Teórica, Atómica y Óptica, Universidad de Valladolid, 47011 Valladolid, Spain}

\date{5th March 2025}

\begin{abstract}
Computing key rates in quantum key distribution (QKD) numerically is essential to unlock more powerful protocols, that use more sophisticated measurement bases or quantum systems of higher dimension. It is a difficult optimization problem, that depends on minimizing a convex non-linear function: the (quantum) relative entropy. Standard conic optimization techniques have for a long time been unable to handle the relative entropy cone, as it is a non-symmetric cone, and the standard algorithms can only handle symmetric ones. Recently, however, a practical algorithm has been discovered for optimizing over non-symmetric cones, including the relative entropy. Here we adapt this algorithm to the problem of computation of key rates, obtaining an efficient technique for lower bounding them. In comparison to previous techniques it has the advantages of flexibility, ease of use, and above all performance.
\end{abstract}
\maketitle

\section{Introduction}
Secret key rates in QKD are usually calculated analytically \cite{gisin2002, scarani2009,xu2020,pirandola2020}. This is only tractable for simple protocols with highly symmetric measurement bases, such as BB84 \cite{bennett1984}, the six-state protocol \cite{bruss1998}, or their generalizations to mutually unbiased bases (MUBs) in higher dimensions \cite{cerf2002,sheridan2010}. Recently there has been intense interest in numerical techniques for the computation of key rates in order to unlock more sophisticated protocols, that use more parameters, arbitrary measurement bases, and higher dimensions.

The best existing techniques are the Gauss-Newton technique from \cite{hu2022}, and the Gauss-Radau technique from \cite{araujo22}. The former consists of an implementation of a bespoke Gauss-Newton solver for the key rate problem, and the latter consists of reducing the problem to a convergent hierarchy of semidefinite programs. Both allow for the computation of optimal key rates for arbitrary protocols, but suffer from several disadvantages. The main one, common to both techniques, is low performance, which limits their applicability to protocols with low dimension. Specific problems of the Gauss-Newton technique are that it doesn't use standard conic optimization techniques, which makes it challenging to implement and inflexible. For example, it cannot handle protocols that use anything other than equality constraints. A specific problem of the Gauss-Radau technique is that for any fixed level of the hierarchy it only delivers an approximation of the key rate, and the cost of achieving a given precision can be prohibitive.

Here we introduce a new technique that is several times faster than both techniques, while also solving their specific issues. Unlike the Gauss-Newton technique, it uses standard conic optimization, making it easy to implement and combine with other kinds of constraints that might be required by the protocols. Unlike the Gauss-Radau technique, it doesn't rely on reformulating the problem as a sequence of semidefinite programs, but instead solves it directly.

We use the framework from \cite{winick2018} to formulate the key rate problem as a convex optimization problem, the minimization of a (quantum) relative entropy over a convex domain. Using the relative entropy cone, which is a non-symmetric cone, it becomes a conic optimization problem. Although for a long time algorithms for optimizing over non-symmetric cones have been known \cite{nesterov1999,tuncel2001,nesterov2012}, they were hardly practical, as they required for instance a tractable barrier function for the dual cone. This changed with the discovery of Skajaa and Ye's algorithm \cite{skajaa2015,papp2017}, and with it solving the key rate problem becomes in principle possible. There is, however, a technical difficulty: this formulation of the key rate problem is often not strictly feasible, and such problems cannot be reliably solved with conic optimization methods. The standard technique to deal with this, called facial reduction \cite{drusvyatskiy2017}, does apply, but the resulting reduced problem is no longer a minimization of the relative entropy. For this reason we introduce a new convex cone to deal directly with the reduced problem. 

We implemented our new cone in the programming language Julia \cite{bezanson2017} as an extension to the solver Hypatia \cite{coey2022}, which implements an improved version of the Skajaa and Ye algorithm \cite{coey2023}. This solver is interfaced by the modeller JuMP \cite{lubin2023}, making it easy to use, and takes advantage of Julia's flexible type system, allowing the user to solve the optimization problems using double, double-double, quadruple, or arbitrary precision.

The paper is organized as follows. In Section \ref{sec:formulation} we reformulate the key rate problem as a conic problem and introduce the QKD cone. In Section \ref{sec:implementation} we propose a barrier function and obtain the derivatives necessary to implement the cone. In Section \ref{sec:examples} we show how to formulate some QKD protocols to compute the key rates with our technique. In Section \ref{sec:numerics} we benchmark our technique against the Gauss-Newton and Gauss-Radau techniques.

\section{QKD rate as a conic problem}\label{sec:formulation}

In quantum key distribution the asymptotic key rate is given by the Devetak-Winter rate \cite{devetak2005}:
\begin{equation}
K \ge H(A|E)-H(A|B),
\end{equation}
where $H(A|E)$ is the conditional von Neumann entropy between Alice and Eve, and $H(A|B)$ the conditional von Neumann entropy between Alice and Bob. Since $H(A|B)$ is completely determined by the measured data, the challenge is to compute $H(A|E)$, which has to be minimized over all quantum states compatible with the measured statistics. Following \cite{winick2018}, it can be expressed in terms of the relative entropy as
\begin{equation}
H(A|E)_{\rho_{ABE}} = D(\mathcal G(\rho_{AB}) \| \mathcal Z(\mathcal G(\rho_{AB}))),
\end{equation}
where $D(\cdot\| \cdot)$ is the relative entropy, $\mathcal G$ is a CP map that takes $\rho_{AB}$ to an extended quantum state that includes the secret key as a subsystem, and $\mathcal Z$ is a CPTP map that decoheres the key subsystem. $\mathcal Z$ is necessarily of the form $\mathcal Z(\rho) = \sum_i \Pi_i \rho \Pi_i$ for some projectors $\Pi_i$ that sum to identity. The optimization problem is thus
\begin{equation}
\begin{gathered}
\min_{\rho}\ D(\mathcal G(\rho) \| \mathcal Z(\mathcal G(\rho))) \\
\text{s.t.}\quad \rho \succeq 0,\quad \tr(\rho) = 1, \\
\tr(E_k \rho) = p_k\ \forall k
\end{gathered}
\end{equation}
where $E_k$ are the POVMs encoding the QKD protocol, and $p_k$ are the estimated probabilities. This formulation makes the computation of key rates a convex optimization problem. To make it a \emph{conic} optimization problem, however, we need to write it in terms of the relative entropy convex cone, namely
\begin{equation}\label{eq:relativeentropycone}
\mathcal K_\text{RE} = \text{cl}\,\{(h,\rho,\sigma) \in \mathbb R \times \mathbb H^{n} \times \mathbb H^{n};\, \rho \succ 0,\  \sigma \succ 0, \  h \ge D(\rho\|\sigma)\},
\end{equation}
where $\mathbb H^{n}$ denotes the space of $n \times n$ complex Hermitian matrices. It becomes then
\begin{equation}
\begin{gathered}
\min_{h,\rho}\ h \\
\text{s.t.}\quad \rho \succeq 0,\quad \tr(\rho) = 1, \\
\tr(E_k \rho) = p_k\ \forall k, \\
\Big(h,\mathcal G(\rho),\mathcal Z(\mathcal G(\rho))\Big) \in \mathcal K_\text{RE}
\end{gathered}
\end{equation}
In principle it can now be solved by using Skajaa and Ye's algorithm, but there's a difficulty: as with any interior-point algorithm, the problem needs to be strictly feasible for the solution to be found reliably and efficiently. In other words, there must exist a point $(h, \rho)$ that satisfies the equality constraints of the conic problem and is in the interior of all cones we use. This can fail here in two places: the equality constraints $\tr(E_k \rho) = p_k$ might be only satisfiable by a quantum state with null eigenvalues, and the map $\mathcal G$ might take a positive definite quantum state to one with null eigenvalues. The map $\mathcal Z$, on the other hand, preserves positive definiteness\footnote{To see that, assume for the sake of contradiction that $\ket{\psi}$ is a null eigenvector of $\mathcal Z(\rho)$ for some positive definite $\rho$. Then $\bra{\psi}\mathcal Z(\rho)\ket{\psi} = \sum_i \bra{\psi}\Pi_i\rho\Pi_i\ket{\psi} = 0$. Since $\rho$ is positive definite, $\bra{\psi}\Pi_i\rho\Pi_i\ket{\psi} > 0$ for all nonzero $\Pi_i\ket{\psi}$. Therefore, to satisfy the equation we must have $\Pi_i\ket{\psi} = 0\ \forall i$. Summing over $i$ and using the fact that $\sum_i \Pi_i = \id$, we arrive at $\ket{\psi} = 0$, contradiction.} \cite{hu2022}.

We note that this same difficulty affects other techniques \cite{winick2018, hu2022, araujo22}: in \cite{winick2018} they perturb the singular matrices with identity to make them full rank, whereas in \cite{hu2022,araujo22} they reformulate the problem in terms of the support of the relevant matrices, a procedure known as facial reduction \cite{drusvyatskiy2017}. Facial reduction is more reliable, elegant, and results in a smaller problem than perturbing with identity, so we adopt this approach here as well.

We would thus like to perform facial reduction along the lines of \cite{hu2022}: first find an isometry $V$ that encodes the support of the feasible $\rho$, such that $\rho = V\sigma V^\dagger$ for full-rank $\sigma$. Then replace the equality constraints with $\tr(V^\dagger E_k V \sigma) =: \tr(F_k \sigma) = p_k$, dropping the redundant ones. Finally, find maps $\gmap$ and $\zmap$ such that $\gmap(\sigma)$ and $\zmap(\sigma)$ are the restrictions of the maps $\sigma \mapsto \mathcal G(V\sigma V^\dagger)$ and $\sigma \mapsto \mathcal Z(\mathcal G(V\sigma V^\dagger))$ to the support of their ranges.

While that's easy to do, the resulting problem is no longer an optimization over the relative entropy; in particular $D(\gmap(\sigma)\|\zmap(\sigma)) \neq D(\mathcal G(\rho) \| \mathcal Z(\mathcal G(\rho)))$. Instead, to obtain the reformulated problem we use the following identity \cite{coles2012}:
\begin{equation}
D(\mathcal G(\rho) \| \mathcal Z(\mathcal G(\rho))) = -H(\mathcal G(\rho)) + H(\mathcal Z(\mathcal G(\rho))),
\end{equation}
where $H(\rho) = - \tr(\rho \log \rho)$ is the von Neumann entropy, together with the simple observation that $H(\mathcal G(\rho)) = H(\gmap(\sigma))$ and $H(\mathcal Z(\mathcal G(\rho))) = H(\zmap(\sigma))$. This motivates the definition of a new convex cone:
\begin{equation}\label{eq:qkdcone}
\mathcal K_\text{QKD}^{\widehat{\mathcal G}, \widehat{\mathcal Z}} = \{(h,\sigma) \in \mathbb R \times \mathbb H^{n};\, \sigma \succeq 0, \  h \ge -H(\widehat{\mathcal G}(\sigma)) + H(\widehat{\mathcal Z}(\sigma)) \}.
\end{equation}
We emphasize that $\gmap$ and $\zmap$ must come from the reduction of $\mathcal G$ and $\mathcal Z$; if we let them be arbitrary CP maps $\mathcal K_\text{QKD}^{\widehat{\mathcal G}, \widehat{\mathcal Z}}$ is not even a convex cone in general.

With the cone in hand, we can finally state the reduced problem:
\begin{equation}\label{eq:conicprogram}
\begin{gathered}
\min_{h,\sigma}\ h \\
\text{s.t.}\quad \tr(\sigma) = 1,\quad \tr(F_k \sigma) = p_k\ \forall k, \\
(h,\sigma) \in \mathcal K_\text{QKD}^{\gmap, \zmap}
\end{gathered}
\end{equation}
Note that the constraint $\rho \succeq 0$ has been dropped; although it's equivalent to $\sigma \succeq 0$, the latter is redundant with the definition of $\mathcal K_\text{QKD}^{\gmap, \zmap}$.

We emphasize that even in the case where no facial reduction is necessary, and $\gmap = \mathcal G$ and $\zmap = \mathcal Z \circ \mathcal G$, it is still advantageous to use the QKD cone \eqref{eq:qkdcone} instead of the relative entropy cone \eqref{eq:relativeentropycone} for four reasons: (i) the constraint $\rho \succeq 0$ can not be dropped in the relative entropy cone unless $\mathcal G$ has a positive inverse, (ii) the QKD cone has one less matrix variable to optimize over, (iii) the derivatives of the barrier function of the QKD cone, that we introduce below, are less computationally demanding, (iv) as the QKD cone explicitly depends on $\gmap$ and $\zmap$, it can exploit their structure; specifically, we use the fact that $\zmap$ necessarily has a block diagonal structure to drastically speed up the computation of several derivatives involving it.

\section{Implementation of the QKD cone}\label{sec:implementation}

In order to successfully optimize over a given convex cone $\mathcal K$, both using Skajaa and Ye's algorithm or traditional methods for symmetric cones, it's essential to provide a logarithmically homogeneous self-concordant barrier (LHSCB) function for it. It is defined as a function $f : \operatorname{int}(\mathcal K) \to \mathbb R$ such that $f(x_i) \to \infty$ along every sequence converging to the boundary of $\mathcal K$, is three times continuously differentiable, strictly convex, and
\begin{gather}
    f(\tau x) = f(x) - \nu \log \tau \quad \forall x \in \operatorname{int}(\mathcal K), \forall \tau > 0,\\
    |\nabla^3 f(x)[\xi,\xi,\xi]| \le 2 (\nabla^2 f(x)[\xi,\xi])^{3/2}\quad \forall x \in \operatorname{int}(\mathcal K), \forall \xi, \label{eq:selfconcordance}
\end{gather}
where $\nabla^2 f(x)[\xi,\xi]$ and $\nabla^3 f(x)[\xi,\xi,\xi]$ denote the Hessian and third-order derivative at point $x$ in direction $\xi$. For an introduction to LHSCB functions and the matrix derivatives see \cite{nesterov2018}.

While most of these properties are straightforward to verify, self-concordance \eqref{eq:selfconcordance} is not, and has only recently been proven for the following barrier function for the relative entropy cone \eqref{eq:relativeentropycone} \cite{fawzi2022}:
\begin{equation}\label{eq:barrierre}
f(h,\rho,\sigma) = - \log(h -D(\rho \| \sigma)) -\operatorname{logdet}(\rho) -\operatorname{logdet}(\sigma).
\end{equation}
We propose as the barrier function for the QKD cone \eqref{eq:qkdcone} its obvious analogue:
\begin{equation}\label{eq:barrierqkd}
f(h,\rho) = - \log(h + H(\gmap(\rho)) - H(\zmap(\rho))) -\operatorname{logdet}(\rho).
\end{equation}
We conjecture that it is also self-concordant based on the similarity and the fact that our extensive numerical tests were successful.

In order to implement the cone \eqref{eq:qkdcone} in the solver Hypatia, we need to provide the gradient, Hessian, and third-order derivative of the barrier function \eqref{eq:barrierqkd}, and choose an initial point \cite{coey2023}. We remark that third-order derivatives are usually not required by primal-dual methods, and in particular not by the original Skajaa-Ye algorithm. Incorporating them is an innovation of Hypatia's algorithm.

\subsection{Gradient, Hessian, and third-order derivative}

To obtain the gradient, Hessian, and third-order derivative of the barrier function \eqref{eq:barrierqkd} we adapt the results of \cite{faybusovich2022} to complex matrices, and combine them with the results of \cite{coey2024} for the relative entropy cone. In the code they are needed in vectorized form, as explained in Appendix \ref{sec:vectorization}. Here we present them in the usual notation for clarity.

Let $g(\rho) = -H(\mathcal{L}(\rho))$ for some positive map $\mathcal{L}$. Then its gradient is
\begin{equation}
\nabla_\rho g(\rho) = \mathcal{L}^\dagger(\id + \log(\mathcal{L}(\rho))),
\end{equation}
where $\mathcal{L}^\dagger$ is the adjoint of $\mathcal{L}$. Its Hessian is the linear operator
\begin{equation}
\nabla^2_{\rho,\rho} g(\rho) = \xi \mapsto \mathcal{L}^\dagger\left(U(\Gamma^{[1]}(\Lambda) \odot (U^\dagger \mathcal{L}(\xi) U))U^\dagger\right),
\end{equation}
where $\odot$ is the elementwise (Schur) product, $\mathcal L(\rho) = U \Lambda U^\dagger$ is a diagonalization of $\mathcal L(\rho)$, and $\Gamma^{[1]}(\Lambda)$ is defined as
\begin{equation}
\Gamma^{[1]}(\Lambda)_{ij} = \begin{cases} \frac{\log(\lambda_i)-\log(\lambda_j)}{\lambda_i-\lambda_j}, & \lambda_i \neq \lambda_j \\ \lambda_i^{-1}, & \lambda_i = \lambda_j \,, \end{cases}
\end{equation}
where $\lambda_i = \Lambda_{ii}$. Its third-order derivative applied to $\xi,\xi$ is
\begin{equation}
\nabla_{\rho,\rho,\rho}^3 g(\rho)[\xi,\xi] = \lmap^\dagger( U_\lmap M_\lmap(\xi) U_\lmap^\dagger)\,,
\end{equation}
where
\begin{equation}\label{D2Log}
    M_{\mathcal{L}}(\xi)_{\,i,j} = 2\sum_k \tilde{\xi}_{i,k} \tilde{\xi}_{k,j} \Gamma^{[2]}_{i,j,k}(\Lambda_\lmap)\,,
\end{equation}
with $\tilde{\xi} = U_{\mathcal L}^\dagger \lmap(\xi) U_{\mathcal L}$ and
\begin{equation}
\Gamma^{[2]}_{i,j,k}(\Lambda) = \begin{cases}
\frac{\Gamma^{[1]}_{i,j}(\Lambda)-\Gamma^{[1]}_{i,k}(\Lambda)}{\lambda_j-\lambda_k}, & \lambda_j \neq \lambda_k \\ 
\frac{\Gamma^{[1]}_{i,j}(\Lambda)-\Gamma^{[1]}_{j,j}(\Lambda)}{\lambda_i-\lambda_j}, & \lambda_i \neq \lambda_j = \lambda_k \\ 
-\lambda_i^{-2}/2, & \lambda_i = \lambda_j = \lambda_k \:.\end{cases}
\end{equation}

The gradient and Hessian of the term $-\operatorname{logdet}(\rho)$ are well-known to be $-\rho^{-1}$ and $\xi \mapsto \rho^{-1}\xi\rho^{-1}$, respectively \cite{nesterov1997}. The third order derivative can be easily calculated to be
\begin{equation}
	(\xi,\zeta) \mapsto - \rho^{-1}\xi\rho^{-1}\zeta\rho^{-1} - \rho^{-1}\zeta\rho^{-1}\xi\rho^{-1}.
\end{equation}

Putting everything together, the gradient of the barrier $f(h,\rho)$ is given by
\begin{align}
\nabla_h f(h,\rho) &= -\frac{1}{u}, \label{eq:gradh}\\
\nabla_\rho f(h,\rho) &= -\frac{1}{u} \nabla_\rho\,u -\rho^{-1}, \label{eq:gradrho}
\end{align}
where $u = h + H(\gmap(\rho)) - H(\zmap(\rho))$ and $\nabla_\rho\,u = -\gmap^\dagger(\id + \log(\gmap(\rho))) + \zmap^\dagger(\id + \log(\zmap(\rho)))$, and the Hessian by
\begin{align}
\nabla_{h,h}^2 f(h,\rho) &= \frac{1}{u^2}, \\
\nabla_{h,\rho}^2 f(h,\rho) &= \frac{1}{u^2} \nabla_\rho\,u, \\
\nabla_{\rho,\rho}^2 f(h,\rho) &= \xi \mapsto \frac{1}{u^2} \tr\big[(\nabla_\rho\,u) \xi\big]\nabla_\rho\,u - \frac{1}{u}\nabla_{\rho,\rho}^2\,u[\xi] + \rho^{-1}\xi\rho^{-1},
\end{align}
where
\begin{equation}
\nabla_{\rho,\rho}^2\,u[\xi] = -\gmap^\dagger\left(U_{\mathcal G}(\Gamma^{[1]}(\Lambda_{\mathcal G}) \odot (U_{\mathcal G}^\dagger \gmap(\xi) U_{\mathcal G}))U_{\mathcal G}^\dagger\right) + \zmap^\dagger\left(U_{\mathcal Z}(\Gamma^{[1]}(\Lambda_{\mathcal Z}) \odot (U_{\mathcal Z}^\dagger \zmap(\xi) U_{\mathcal Z}))U_{\mathcal Z}^\dagger\right),
\end{equation}
and $\gmap(\rho) = U_{\mathcal G} \Lambda_{\mathcal G} U_{\mathcal G}^\dagger$ and $\zmap(\rho) = U_{\mathcal Z} \Lambda_{\mathcal Z} U_{\mathcal Z}^\dagger$ are diagonalizations of $\gmap(\rho)$ and $\zmap(\rho)$.

The third order derivatives are given by:
\begin{align}
	\nabla_{h,h,h}^3 f(h,\rho) &= -\frac{2}{u^3},  
 \\
	\nabla_{h,h,\rho}^3 f(h,\rho) &= -\frac{2}{u^3}\nabla_\rho u, 
 \\
	\nabla_{h,\rho,\rho}^3 f(h,\rho) &= \xi \mapsto -\frac{2}{u^3} \tr\big[(\nabla_\rho\, u) \xi\big]\nabla_\rho u + \frac{1}{u^2}\nabla_{\rho,\rho}^2\,u[\xi],
 \\
	\nabla_{\rho,\rho,\rho}^3 f(h,\rho) &= (\xi,\zeta) \mapsto 
    - \frac{2}{u^3}  \tr\big[(\nabla_\rho\, u) \xi \big] \tr\big[(\nabla_\rho\, u) \zeta\big]\nabla_\rho u \nonumber 
    + \frac{1}{u^2}\tr\big[(\nabla_{\rho,\rho}^2\, u[\zeta]) \xi\big] \nabla_\rho u \nonumber 
    \\	&\qquad\qquad\quad\, 
    + \frac{1}{u^2}\tr\big[(\nabla_\rho\, u) \xi\big] \nabla_{\rho,\rho}^2 u[\zeta] \nonumber 
    + \frac{1}{u^2}\tr\big[(\nabla_\rho\, u) \zeta\big] \nabla_{\rho,\rho}^2 u[\xi] \nonumber 
    \\	&\qquad\qquad\quad\, 
    - \frac{1}{u} \nabla_{\rho,\rho,\rho}^3 u[\xi,\zeta]
    - (\rho^{-1}\xi\rho^{-1}\zeta\rho^{-1} + \rho^{-1}\zeta\rho^{-1}\xi\rho^{-1}).
\end{align}
In particular, applying $\nabla_{\rho,\rho,\rho}^3 f(h,\rho)$ to the point $[\xi,\xi]$ gives
\begin{align}
	\nabla_{\rho,\rho,\rho}^3 f(h,\rho)[\xi,\xi] &= - \frac{2}{u^3}  \big(\tr\big[(\nabla_\rho\, u) \xi \big]\big)^2\nabla_\rho u + \frac{1}{u^2}\tr\big[(\nabla_{\rho,\rho}^2\, u[\xi]) \xi\big] \nabla_\rho u \nonumber \\
	&\quad + \frac{2}{u^2}\tr\big[(\nabla_\rho\, u) \xi\big] \nabla_{\rho,\rho}^2 u[\xi] - \frac{1}{u} \nabla_{\rho,\rho,\rho}^3 u[\xi,\xi] \nonumber \\
	&\quad - 2\rho^{-1}\xi\rho^{-1}\xi\rho^{-1},
\end{align}
where
\begin{equation}
\nabla_{\rho,\rho,\rho}^3 u[\xi,\xi] = - {\gmap}^\dagger( U_{\g} M_{\g}(\xi) U_{\g}^\dagger) + \zmap^\dagger( U_{\z} M_{\z}(\xi) U_{\z}^\dagger),
\end{equation}  
and $M_\g(\xi)$ and $M_\z(\xi)$ are the analogues of \eqref{D2Log}.

\subsection{Initial point}
The initial point can in principle be chosen to be any point in the interior of the cone \cite{skajaa2015}, but following \cite{coey2023,dahl2022} we'd like to use the central point, as it provides better performance. It is defined as the minimizer of
\begin{equation}
     \min\limits_{h,\rho} f(h,\rho)+\frac{1}{2}\| (h,\rho) \|^2_2 \,,
\end{equation}
where $f(h,\rho)$ is our barrier function \eqref{eq:barrierqkd}, and $\| (h,\rho) \|^2_2 = h^2 + \langle \rho, \rho \rangle$. The minimizer is obtained by taking the gradient of the objective and setting it to zero:
\begin{align}
(h,\rho) = -\nabla f(h,\rho).
\end{align}
This results in the pair of equations
\begin{align}
    h  & = \frac{1}{u},\\
    \rho  &= \frac{1}{u} \nabla_\rho\,u +\rho^{-1},
\end{align}
where we are using the gradient from equations \eqref{eq:gradh}--\eqref{eq:gradrho}.

Let now $u = h - D$, where $D = -H(\gmap(\rho)) + H(\zmap(\rho))$. We get then\footnote{The negative solution for $h$ is excluded as it is outside the cone.}
\begin{gather}
    h=\frac{D}{2} + \sqrt{1+\frac{D^2}{4}}, \label{eq:h}\\
\rho - \rho^{-1}= h \left(-\gmap^\dagger(\id + \log(\gmap(\rho))) + \zmap^\dagger(\id + \log(\zmap(\rho)))\right). \label{eq:rho}
\end{gather}
While we have been unable to obtain a general solution of these equations, we note that often the maps satisfy $\gmap^\dagger(\id + \log(\gmap(\id))) = \zmap^\dagger(\id + \log(\zmap(\id)))$. In this case $\rho = \id$ will satisfy equation \eqref{eq:rho} for any value of $h$, which we can then choose to be given by equation \eqref{eq:h}. We use therefore these values as our initial point.

Note that even when equation \eqref{eq:rho} is not satisfied this point is still in the interior of the cone, as $\id$ is full rank and $\frac{D}{2} + \sqrt{1+\frac{D^2}{4}} > D$, and thus is a valid starting point.

\section{Examples}\label{sec:examples}

\subsection{BB84}\label{sec:bb84}

In order to illustrate the technique let us start with a toy example, the computation of the key rate for the entanglement based version of the BB84 protocol, using as estimated parameters the qubit error rates (QBERs) in the $X$ and $Z$ bases $q_x$ and $q_z$. If we use only the $Z$ basis to generate key, the key map $\mathcal G$ can be taken to be identity, and the decoherence map $\mathcal Z$ is $\rho \mapsto \sum_{i=0}^1 (\proj{i} \otimes \id) \rho (\proj{i} \otimes \id)$. The analytical expression for $H(A|E)$ is $1-h(q_x)$, where $h$ is the binary entropy. 

In the case where $(q_x,q_z) \in (0,1)^{\times 2}$ there is a full rank state compatible with the constraints and the support of the range of $\g$ and $\z$ is the full space, so no facial reduction is needed. Then $\gmap = \g$ and $\zmap = \z \circ \g$, and the conic program is
\begin{equation}
\begin{gathered}
\min_{h,\rho}\ h \\
\text{s.t.}\quad \tr(\rho) = 1,\quad \tr(Q_x \rho) = q_x, \quad \tr(Q_z \rho) = q_z, \\
(h,\rho) \in \mathcal K_\text{QKD}^{\gmap, \zmap},
\end{gathered}
\end{equation}
where $Q_x$ and $Q_z$ are the projectors that produce the QBERs.

In the case where $q_z = 0$ and $q_x \in (0,1)$ the support of the feasible $\rho$ is $\operatorname{span}\{\ket{\phi^+},\ket{\phi^-}\}$. Letting $V = \ketbra{\phi^+}{0} + \ketbra{\phi^-}{1}$ we can write $\rho = V\sigma V^\dagger$ for a $2 \times 2$ matrix $\sigma$. The constraint $\tr(V^\dagger Q_z V \sigma) = 0$ becomes tautological and is dropped, and the constraint $\tr(V^\dagger Q_x V \sigma) = q_x$ is reduced to $\tr(\proj{1}\sigma) = q_x$. The map $\sigma \mapsto V\sigma V^\dagger$ has range supported on the range of $V$, namely $\operatorname{span}\{\ket{\phi^+}, \ket{\phi^-}\}$, so to restrict it there we simply remove the isometry, obtaining $\gmap(\sigma) = \sigma$. The map $\sigma \mapsto \z(V\sigma V^\dagger)$ has range supported on $\operatorname{span}\{\ket{00}, \ket{11}\}$, so we restrict it there, choosing the Kraus operators of $\zmap$ to be $\{\ketbra{0}{+}, \ketbra{1}{-}\}$.

In the case where $q_x = 0$ and $q_z \in (0,1)$ the support of the feasible $\rho$ is $\operatorname{span}\{\ket{\phi^+},\ket{\psi^+}\}$. Letting $V = \ketbra{\phi^+}{0} + \ketbra{\psi^+}{1}$ we can write $\rho = V\sigma V^\dagger$ for a $2 \times 2$ matrix $\sigma$. The constraint $\tr(V^\dagger Q_x V \sigma) = 0$ becomes tautological and is dropped, and the constraint $\tr(V^\dagger Q_z V \sigma) = q_z$ is reduced to $\tr(\proj{1}\sigma) = q_z$. To reduce the map $\sigma \mapsto V\sigma V^\dagger$ we again simply remove the isometry, obtaining $\gmap(\sigma) = \sigma$. The map $\sigma \mapsto \z(V\sigma V^\dagger)$ this time has full rank range, so its reduction is simply itself, $\zmap(\sigma) = \z(V\sigma V^\dagger)$.

In the case where $q_x = q_z = 0$ the only feasible $\rho$ is $\proj{\phi^+}$ so there's nothing to optimize.

\subsection{DMCV}\label{sec:dmcv}

A more interesting case of facial reduction is when a POVM $\{E_i\}_{i=0}^{n-1}$ is used to generate key. This is the case for example in discrete-modulated continuous variable (DMCV) QKD \cite{lin2019}. For simplicity, assume that the estimated probabilities are compatible with a full-rank state, so that facial reduction is only needed for the maps $\g$ and $\z$.

Let $V = \sum_{i=0}^{n-1} \id_A \otimes \sqrt{E_i}_B \otimes \ket{i}_a $ be the standard Naimark dilation of the POVM\footnote{Note that here we are applying the POVMs on Bob's side, as is standard in DMCV.}, where $a$ is an ancilla subsystem. Then $\g(\rho) = V\rho V^\dagger$, and $\z$ has Kraus operators $\{\id_{AB} \otimes \proj{i}_a\}_{i=0}^{n-1}$. As before, since $\g$ is just an isometry, the reduction consists simply of removing it, and $\gmap(\rho) = \rho$. The Kraus operators of $\z \circ \g$ are $\{\id_A \otimes \sqrt{E_i}_B \otimes \ket{i}_a\}_{i=0}^{n-1}$. It is easy to see that if all $E_i$ are full rank then $\z \circ \g$ has full rank range, and $\zmap = \z \circ \g$. Otherwise further reduction is needed (or starting with a more appropriate Naimark dilation).

For the purpose of benchmarking we implemented here the heterodyne protocol from \cite{lin2019} (a more sophisticated application of our technique to DMCV is presented in \cite{pascualgarcia2024}). In it the POVMs are in fact full rank and the facial reduction is as described in the previous paragraph. The main parameter of the protocol is the number of photons at which the Fock space is cut off, $N_c$. The quantum state $\rho$ has dimension $4(N_c+1)$, and the Kraus operators of $\zmap$ are $16(N_c+1) \times 4(N_c + 1)$. As it is a prepare-and-measure protocol, it also has constraints on the partial trace of $\rho$, which in general may introduce null eigenvalues and necessitate further facial reduction. Here it is not the case.

\subsection{MUB protocol}\label{sec:mub}

In the MUB protocol introduced in \cite{cerf2002,sheridan2010}, Alice and Bob measure a complete set of MUBs for some prime $d$, with Bob's bases being the transpose of Alice's, and estimate the probability of getting equal results. As in \cite{araujo22}, here there is no such limitations of using prime dimensions, equal results, or even exact MUBs.

As it is an entanglement-based protocol and we are generating key in a single basis, the key map $\g$ can be taken to be identity, and the decoherence map $\z$ to have Kraus operators $\{\proj{i} \otimes \id\}_{i=0}^{d-1}$. When the estimated statistics are compatible with a full-rank state the facial reduction is trivial: $\gmap = \g$ and $\zmap = \z \circ \g$.

\subsection{Overlapping bases protocol}\label{sec:overlap}

In the overlapping bases protocol introduced in \cite{araujo22}, Alice and Bob measure only nearest-neighbour superpositions, with again Bob's bases being the transpose of Alice's. $\g$ and $\z$ are the same as in the MUB protocol, and the facial reduction is also trivial the estimated statistics are compatible with a full-rank state: $\gmap = \g$ and $\zmap = \z \circ \g$.

We note, however, that in \cite{araujo22} the protocol was restricted to even $d \ge 2$, whereas here we use it also for odd $d$.

\subsection{Dealing with experimental data}

As argued in \cite{araujo22}, in order to deal with experimental data one cannot naïvely set the probabilities $\mathbf{p}$ to be equal to the measured relative frequencies $\mathbf{f}$. Instead, one should estimate the covariance matrix $\Sigma$, and minimize $H(A|E)$ over the desired confidence region, approximated as the intersection of the ellipsoid $\big\|\Sigma^{-\frac12}(\mathbf{p}-\mathbf{f})\big\|_2 \le \chi$ with the set of parameters corresponding to valid quantum states.

One can do that with a simple modification of our conic program \eqref{eq:conicprogram}:
\begin{equation}\label{eq:conicprogramexperimental}
\begin{gathered}
\min_{h,\sigma,\mathbf{p}}\ h \\
\text{s.t.}\quad \tr(\sigma) = 1,\quad \tr(\mathbf{F} \sigma) = \mathbf{p}, \quad \big\|\Sigma^{-\frac12}(\mathbf{p}-\mathbf{f})\big\|_2 \le \chi, \\
(h,\sigma) \in \mathcal K_\text{QKD}^{\gmap, \zmap}
\end{gathered}
\end{equation}
where the constraint we added is a second-order-cone constraint, which is supported by almost every conic solver in existence, and in particular the one we are using, Hypatia.

\section{Numerical results}\label{sec:numerics}

\subsection{Performance}

In order to illustrate the performance of our technique, we ran the conic program \eqref{eq:conicprogram} and compared the running time to the Gauss-Newton technique \cite{hu2022} and the Gauss-Radau technique \cite{araujo22}. We didn't compare to the Frank-Wolfe technique from \cite{winick2018} because \cite{hu2022} already demonstrated superior performance. We didn't compare to the technique from \cite{karimi2024} because they only use the vanilla relative entropy cone; as such they cannot perform facial reduction and are unable to handle the QKD problem in full generality.

The protocols we benchmarked are the DMCV, MUB and overlap protocols described in subsections \ref{sec:dmcv}, \ref{sec:mub} and \ref{sec:overlap}. We chose them because they are currently of high experimental and theoretical interest \cite{bulla2023,lib24,pascualgarcia2024,primaatmaja2024}, and because they have a free parameter controlling the dimension that allows us to see how the solution time scales.

For the DMCV protocol we used the same parameters as in \cite{hu2022}: amplitude of the coherent states $\alpha=0.35$, noise $\xi=0.05$, distance $L = 60$, and transmittance $\eta = 10^{-0.02 L}$. For the MUB and overlap protocols we used the probabilities from the isotropic state $\rho_\text{iso}(v) = v \proj{\phi^+} + (1-v) \id/d^2$ with visibility $v=0.95$. For the MUB protocol we computed the probabilities using all bases, included the complex ones, whereas for the overlap protocol we restricted ourselves to the real ones. We did this in order to illustrate an additional advantage of our technique: since it uses standard conic optimization, we can avail ourselves of standard symmetrization techniques to show that we can optimize over real variables only, as done in \cite{araujo22}. This provides an additional boost in performance. 

The calculations were done on an AMD Ryzen Threadripper Pro 5955WX with 4 GHz and 16 cores, on a machine with 512 GiB RAM running Ubuntu Linux 22.04. Our code was run with Julia 1.11.1 and the modeller JuMP \cite{lubin2023}, and the two other techniques with MATLAB 2023b. For the Gauss-Radau technique we used the solver MOSEK 10.1 \cite{mosek} and the modeller YALMIP \cite{yalmip}. In all cases we have only reported the time taken by the solver, discounting the time taken to set up the problem, and in the Gauss-Newton case the time taken to do facial reduction numerically.

\begin{figure}[h!]
	\centering
 	\begin{tikzpicture}
		\begin{axis}[%
			scale only axis,
            ymode = log,
			grid=major,
			xlabel={$d$},
            ylabel = {time (s)},
			axis background/.style={fill=white},
			legend style={at={(0.94,0.3)},legend cell align=left, align=left, draw=white!15!black}
			]
            \addplot table[col sep=space] {plot_data/overlap_ns};
        	\addlegendentry{QKD cone}
            \addplot table[col sep=space] {plot_data/overlap_gn};
        	\addlegendentry{Gauss-Newton}   
            \addplot[mark=triangle*, teal] table[col sep=space] {plot_data/overlap_gr};
			\addlegendentry{Gauss-Radau}   
        \end{axis}
	\end{tikzpicture}
	\caption{Running time in seconds (logarithmic scale) as a function of the local state dimension for the overlap protocol.}
 \label{fig:overlap_times}
\end{figure}
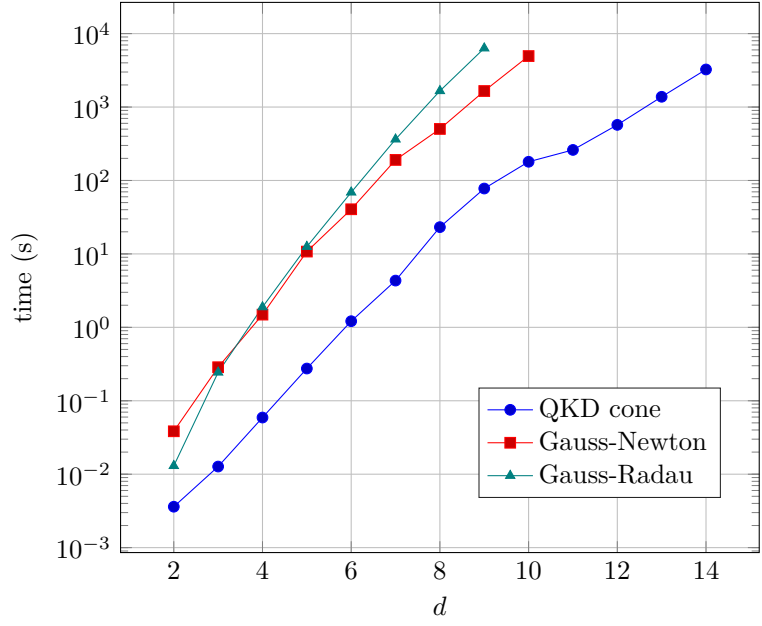

The results are shown in Figures \ref{fig:dmcv_times}, \ref{fig:mub_times}, and \ref{fig:overlap_times}.

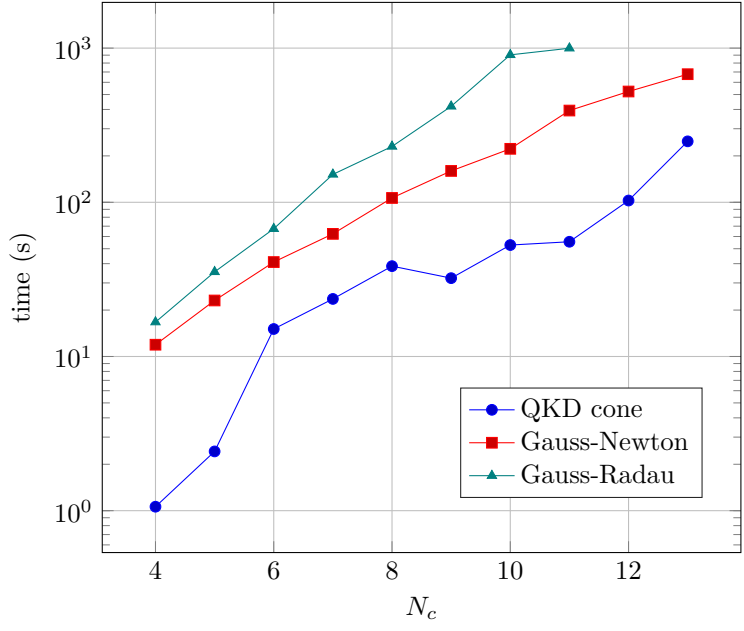
\begin{figure}[h!]
	\centering
 	\begin{tikzpicture}
		\begin{axis}[%
			scale only axis,
            ymode = log,
			grid=major,
			xlabel={$N_c$},
            ylabel = {time (s)},
			axis background/.style={fill=white},
			legend style={at={(0.94,0.3)},legend cell align=left, align=left, draw=white!15!black}
			]
            \addplot table[col sep=space] {plot_data/dmcv_ns};
        	\addlegendentry{QKD cone}
            \addplot table[col sep=space] {plot_data/dmcv_gn};
        	\addlegendentry{Gauss-Newton}   
            \addplot[mark=triangle*, teal] table[col sep=space] {plot_data/dmcv_gr};
			\addlegendentry{Gauss-Radau}   
        \end{axis}
	\end{tikzpicture}
	\caption{Running time in seconds (logarithmic scale) as a function of the photon cut-off number for the DMCV protocol.}
 \label{fig:dmcv_times}
\end{figure}  
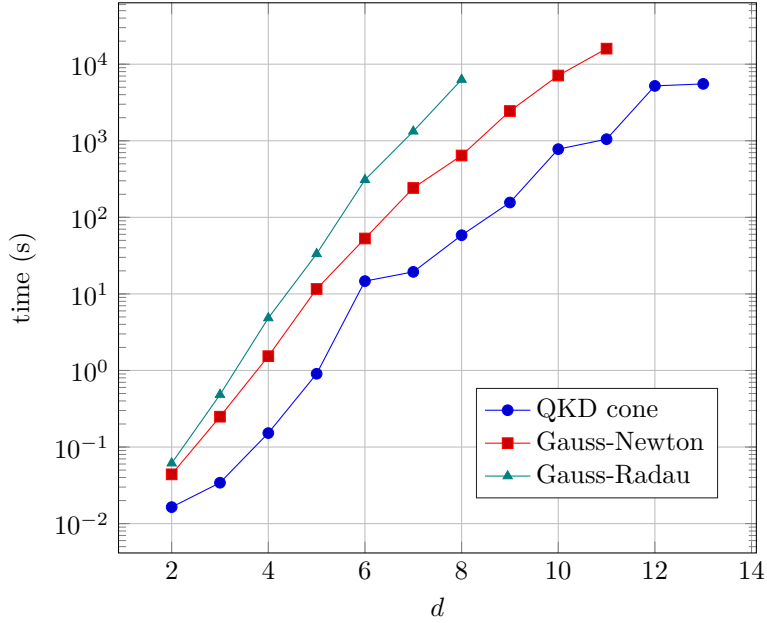
\begin{figure}[h!]
	\centering
	\begin{tikzpicture}
		\begin{axis}[%
			scale only axis,
            ymode = log,
			grid=major,
			xlabel={$d$},
            ylabel = {time (s)},
			axis background/.style={fill=white},
			legend style={at={(0.94,0.3)},legend cell align=left, align=left, draw=white!15!black}
			]
            \addplot table[col sep=space] {plot_data/mub_ns};
        	\addlegendentry{QKD cone}   
            \addplot table[col sep=space] {plot_data/mub_gn};
        	\addlegendentry{Gauss-Newton}   
            \addplot[mark=triangle*, teal] table[col sep=space] {plot_data/mub_gr};
			\addlegendentry{Gauss-Radau}   
        \end{axis}
	\end{tikzpicture}
	\caption{Running time in seconds (logarithmic scale) as a function of the local state dimension for the MUB protocol. Note that for $d=6, 10$, and $12$ the bases used are only roughly unbiased.}
 \label{fig:mub_times}
\end{figure}

\subsection{Precision}

In order to obtain results with higher precision, one might try setting tighter tolerances for the conic solver. This is however not fruitful, as very quickly one meets fundamental obstacles due to the limited precision with which the standard 64-bit floating point numbers can perform computations or even represent the problem parameters. For this reason we have used instead different implementations of floating point numbers that provide more bits of precision. This is easy due to Julia's flexible type system, which is fully taken advantage of by the solver Hypatia and the modeller JuMP; in order to use a different type we simply need to give it as a parameter. Hypatia automatically sets the appropriate tolerances for each type. We have used the following four types:
\begin{itemize}
    \item[\texttt{Float64}] Default 64-bit floating point number implementing the IEEE 754 standard, known as double. Has 53 bits of precision.
    \item[\texttt{Float128}] 128-bit floating point number implementing the IEEE 754 standard, known as quadruple. Has 113 bits of precision. Provided by the package Quadmath \cite{quadmath2024}, which wraps the GCC library libquadmath.
    \item[\texttt{Double64}] Non-standard implementation of a 128-bit floating point number as a pair of 64-bit floating point numbers, known as double-double. Much faster than \texttt{Float128}, but has smaller precision and smaller range of exponent. Has 106 bits of precision. Provided by the package DoubleFloats \cite{sarnoff2024}.
    \item[\texttt{BigFloat}] Arbitrary precision floating-point number. We used the default settings, with which it occupies 640 bits of memory and provides 256 bits of precision. Provided by the Julia standard library, wrapping the GNU MPFR library \cite{fousse2007}.
\end{itemize}
In order to evaluate the error for which type, we solved the conic program \eqref{eq:conicprogram} for the protocols we know the analytical answers: BB84 \ref{sec:bb84} and MUB \ref{sec:mub}. The results are shown in Table \ref{tab:precision}. 

Although the analytical answer is also known for the DMCV protocol \ref{sec:dmcv} for the case $\xi=0$, a comparison is not meaningful: after facial reduction there is a unique state compatible with the constraints, so there's nothing to optimize. We have nevertheless provided the analytical answer and the facial reduction in the accompanying source code for the interested reader.

\begin{table}[h!]
    \centering
    \begin{tabular}{c|cc}
         & BB84 & MUB\\ \hline
        \texttt{Float64} & $5.4\times 10^{-8}$ & $5.7\times 10^{-8}$ \\
        \texttt{Double64} & $1.4\times10^{-13}$ & $1.6\times 10^{-12}$ \\
        \texttt{Float128} & $1.7\times10^{-14}$ & $7.7\times 10^{-14}$ \\
        \texttt{BigFloat} & $7.9\times10^{-32}$ & $1.6\times 10^{-31}$\\
    \end{tabular}
    \caption{Absolute difference between analytical $H(A|E)$ and the one computed via the conic program \eqref{eq:conicprogram} for various floating point implementations, for the BB84 and MUB protocols. For the BB84 protocol we used the parameters $q_x=q_z = 0.025$, and for the MUB protocol we used dimension 3 with 4 bases and the probabilities of an isotropic state with visibility $v=0.95$.}
    \label{tab:precision}
\end{table}

Note that in general a number that has a finite decimal expansion does not have a finite binary expansion. For example, $0.95_{10} = 0.1111\overline{0011}_2$. Therefore, even when giving input parameters that at first sight seem exact, one often incurs in truncation errors. Therefore, in order to take advantage of higher precision types, the input parameters also need to use them\footnote{One must also be careful to write $0.95$ as (e.g.) \texttt{Double64(95)/100} instead of \texttt{Double64(0.95)}, as the latter first computes $0.95$ as a \texttt{Float64} before converting it to \texttt{Double64}.}. For this reason our example codes don't require the type desired for the computation to be specified, but rather read it from the type of the input parameters.

It's important to emphasize that the increased precision comes at the cost of a much longer running time. Not only the elementary operations take longer, but also the number of iterations increases in order to meet the tighter tolerances.

\section{Conclusion}

In this work we have introduced a new technique for obtaining the asymptotic key rate in quantum key distribution. It is based on defining a new convex cone for the key rate problem and optimizing over it with standard conic optimization methods. It provides a dramatic improvement in performance with respect to previous techniques. Moreover, it is flexible and easy to use, as one can freely combine it with other convex cones, and it is interfaced through a convenient modelling language, JuMP.

As a primal-dual method, it directly provides a lower bound through the value of the dual objective. However, we lack an explicit expression for the dual cone, which stops us from converting the dual minimizer into an independent witness for the lower bound. One can nevertheless construct such a witness from the primal minimizer as described in \cite{winick2018} and done for example in \cite{pascualgarcia2024}.

The most pressing issue is to adapt the method to finding finite key rates. Although one can use the asymptotic key rate to obtain a finite key rate via the (generalized) entropy accumulation theorem \cite{dupuis2020,metger2022}, as done for example in \cite{pascualgarcia2024}, such bounds are known to be loose. Tight bounds are provided by more recent techniques \cite{vanhimbeeck2024,arqand2024}, that however require optimization over Rényi entropies. It would be interesting to apply conic methods to this case.

\section{Code availability}

The implementation of the cone introduced in this paper and the code for the examples shown are available in \url{https://github.com/araujoms/ConicQKD.jl}.

\section{Related work}

While conducting this research, we found that He et al. \cite{he2024} had independently applied the Skajaa and Ye algorithm to the problem of QKD using a specialized cone, and proved the self-concordance of the barrier function \eqref{eq:barrierqkd}. In version 0.2 of our code we implemented the technique for inverting the Hessian described in their Appendix B.2.

\section{Acknowledgements}

The research of A.G.L., P.V.P., M.C.C. and M.A. was supported by the European Union--Next Generation UE/MICIU/Plan de Recuperación, Transformación y Resiliencia/Junta de Castilla y León. M.A was also supported by the Spanish Agencia Estatal de Investigación, Grant No. RYC2023-044074-I. We thank Chris Coey and Lea Kapelevich for help with Hypatia's code and useful discussions.


\printbibliography

\appendix

\section{Vectorization}\label{sec:vectorization}

Let $\vec : \mathbb C^{d_O\times d_I} \to \mathbb C^{d_O d_I}$ be the col-major vectorization of a real or complex rectangular matrix. It is useful to represent it as
\begin{equation}
\vec(X) = \id_{d_I} \otimes X \vec(\id_{d_I}),
\end{equation}
where $\vec(\id_{d_I}) = \sum_{i=0}^{d_I-1} \ket{ii}$. It is common to write $\vec(X)$ as $\Ket{X}$. A useful identity is
\begin{equation}\label{eq:vecidentity}
\vec(ABC) = C^T \otimes A \vec(B).
\end{equation}
For implementation purposes we work with a non-redundant vectorization for Hermitian matrices $\svec(X)$. The real and complex cases differ. In the real case, $\svec : \mathbb R^{d \times d} \to \mathbb R^{d(d+1)/2}$ is the col-major vectorization of the upper triangle, with the off-diagonals multiplied by $\sqrt{2}$. In the complex case, $\svec : \mathbb C^{d \times d} \to \mathbb R^{d^2}$ additionally splits the complex numbers into the real part and minus the imaginary part\footnote{This is an arbitrary choice, made to coincide with the code.}, storing them consecutively.

The factor of $\sqrt{2}$ multiplying the off-diagonals is necessary to ensure that
\begin{equation}
\langle X, Y \rangle = \langle \vec(X), \vec(Y) \rangle = \langle \svec(X), \svec(Y) \rangle\quad\forall X,Y.
\end{equation}
Let $V$ be the isometry\footnote{This is an abuse of notation since the operator is different in the real and complex cases. Also note that in the complex case $V$ is additionally unitary.} such that
\begin{equation}
\vec(X) = V\svec(X)
\end{equation}
for all Hermitian $X$. Suppose you are interested in representing a linear function $f$ as a matrix $F$ such that
\begin{equation}
F\vec(X) = \vec(f(X)) \quad \forall X.
\end{equation}
This is easy to do using identity \eqref{eq:vecidentity} and linearity. If this function is additionally Hermitian-preserving we have that
\begin{equation}
V^\dagger F V \svec(X) = \svec(f(X))
\end{equation}
for all Hermitian $X$.

This is mainly useful for proving theorems, as a direct computation of $V^\dagger F V$ via this formula is inefficient. For the particular case of $f(X) = KXK^\dagger$ we implemented an efficient function to compute it, $\operatorname{skron}(K) = V^\dagger (\bar{K} \otimes K) V$.
\end{document}